\documentclass[3p,times,procedia]{elsarticle}
\usepackage{nupha_ecrc}

\volume{00}

\firstpage{1}

\journalname{Nuclear Physics A}

\runauth{B. Duclou\'e et al.}

\jid{nupha}

\jnltitlelogo{Nuclear Physics A}

\usepackage{amssymb}

\usepackage{amsmath}
\usepackage{bm}
\usepackage{hyperref}
\usepackage[utf8]{inputenc}

\usepackage[figuresright]{rotating}

\newcommand{\nn}{\nonumber\\}
\newcommand{\dif}{{\rm d}}
\newcommand{\rme}{{\rm e}}
\newcommand{\rmi}{{\rm i}}
\newcommand{\rmtr}{{\rm tr}}
\newcommand{\bk}{\bm{k}}
\newcommand{\bx}{\bm{x}}
\newcommand{\by}{\bm{y}}
\newcommand{\br}{\bm{r}}
\newcommand{\abar}{\bar{\alpha}_s}

\begin{document}

\begin{frontmatter}



\dochead{XXVIIth International Conference on Ultrarelativistic Nucleus-Nucleus Collisions\\ (Quark Matter 2018)}

\title{On the use of a running coupling in the calculation of forward hadron production at next-to-leading order}

\author[cea]{B.~Duclou\'e}
\author[cea]{E.~Iancu}
\author[jyu]{T.~Lappi}
\author[columbia]{A.H.~Mueller}
\author[cea]{G.~Soyez}
\author[ect]{D.N.~Triantafyllopoulos}
\author[tum]{Y.~Zhu}

\address[cea]{Institut de Physique Th\'{e}orique, Universit\'{e} Paris-Saclay, CNRS, CEA, F-91191 Gif-sur-Yvette, France}
\address[jyu]{Department of Physics, 40014 University of Jyv\"{a}skyl\"{a}, Finland and Helsinki Institute of Physics, 00014 University of Helsinki, Finland}
\address[columbia]{Department of Physics, Columbia University, New York, NY 10027, USA}
\address[ect]{European Centre for Theoretical Studies in Nuclear Physics and Related Areas (ECT*) and Fondazione Bruno Kessler, Strada delle Tabarelle 286, I-38123 Villazzano (TN), Italy}
\address[tum]{Physik-Department, Technische Universit{\" a}t M{\" u}nchen, D-85748 Garching, Germany}

\begin{abstract}
We study a puzzle raised recently regarding the running coupling prescription used in the calculation of forward particle production in proton-nucleus collisions at next-to-leading order: using a coordinate space prescription which is consistent with the one used in the high energy evolution of the target leads to results which can be two orders of magnitude larger than the ones obtained with a momentum space prescription. We show that this is an artefact of the Fourier transform involved when passing between coordinate and momentum space and propose a new coordinate space prescription which avoids this problem.
\end{abstract}

\begin{keyword}
Color Glass Condensate \sep Saturation
\end{keyword}

\end{frontmatter}


\section{Introduction}

Forward particle production in high energy proton-proton or proton-nucleus collisions is an important probe of the nuclear wavefunction at small $x$, where nonlinear effects such as gluon saturation are expected to become sizable. In the past few years, the Color Glass Condensate (CGC) effective theory, which is the natural framework to study such processes, was promoted to next-to-leading order (NLO) accuracy, as required to improve the predictability of this formalism. This includes the NLO corrections both to the Balitsky-Kovchegov (BK) evolution~\cite{Balitsky:1995ub,Kovchegov:1999yj} of gluon densities in a dense nuclear target and to the hard part describing the coupling of a dilute projectile with this target. Unfortunately, the first numerical studies implementing these corrections met with unphysical results such as instability of the NLO high energy evolution~\cite{Lappi:2015fma} and negativity of the forward particle production cross-section~\cite{Stasto:2013cha}. In the latter case, the problem appears in the range of semi-hard transverse momenta where the CGC formalism is supposed to be applicable. As shown in~\cite{Ducloue:2016shw}, this issue is related to the way one separates the target evolution from the impact factor. In~\cite{Iancu:2016vyg}, a reformulation of the cross-section was proposed, which leads to positive results at all transverse momenta at fixed coupling as demonstrated explicitly in~\cite{Ducloue:2017mpb} (a similar observation was recently made in the calculation of the DIS structure functions at NLO~\cite{Ducloue:2017ftk}). However, the correct way of implementing the running of the coupling is still an issue. Indeed, the BK equation is most naturally solved in coordinate space, while the cross-section is written in momentum space. In~\cite{Ducloue:2017mpb} it was shown that a mixed treatment, where different prescriptions for the coupling are used in the impact factor and when solving the BK equation, can make the negativity issue appear again. On the other hand, it was found that the use of the same coordinate space prescription in the whole calculation leads to another problem, with the NLO result becoming orders of magnitude larger than the LO one at large transverse momenta. Our main goal here is thus to understand the origin of this puzzle and to identify a running coupling prescription that can lead to physical results.

\section{Results}

For simplicity, we focus here on the $q \to q$ channel and do not consider the fragmentation functions. The LO quark multiplicity reads
\begin{equation}
\label{lonc}
\frac{\dif N^{{\rm LO}}}{\dif^2\bk\, \dif \eta} 
=\, \frac{x_p q(x_p)}{(2\pi)^2}\,
\mathcal{S}(\bk,X_g)\,,
\end{equation}
where $\bk$ and $\eta$ are the transverse momentum and rapidity of the produced quark respectively, $x_p=(k_\perp e^\eta)/\sqrt{s}$, $X_g=(k_\perp e^{-\eta})/\sqrt{s}$, and $q(x)$ is the quark distribution in the projectile proton. $\mathcal{S}$ is the Fourier transform of the dipole correlator in the color field of the target,
\begin{equation}
\mathcal{S}(\bk,X) = \int \dif^2\br\, 
\rme^{-\rmi \bk \cdot \br}
S(\br, X) \,, \quad S(\bx,\by;X) = \frac{1}{N_c}
\left\langle \rmtr \left[V(\bx) V^{\dagger}(\by) \right]\right\rangle_{X}\,,
\end{equation}
and its evolution as a function of $X$ obeys the Balitsky-Kovchegov equation. The LO multiplicity~(\ref{lonc}) receives NLO corrections proportional to the $N_c$ and $C_F$ color factors which have been computed in~\cite{Chirilli:2011km,Chirilli:2012jd}. Let us first consider the $N_c$ terms which were identified in~\cite{Ducloue:2016shw} as the origin of the negativity problem observed in~\cite{Stasto:2013cha}. The sum of the LO and $N_c$ NLO contributions reads, in the ``unsubtracted'' form~\cite{Iancu:2016vyg},
\begin{align}
\frac{\dif N^{{\rm LO} + N_c}}{\dif^2\bk\, \dif \eta} 
=\, &
\frac{x_p q(x_p)}{(2\pi)^2}
\mathcal{S}(\bk,X_0)	+
\frac{1}{4\pi}\int_0^{1-X_g/X_0} \dif \xi\,
\frac{1+\xi^2}{1-\xi}
\nn
& \times \left[\Theta(\xi-x_p)\frac{x_p}{\xi}\,
q\left(\frac{x_p}{\xi}\right)
\mathcal{J}(\bk,\xi,X(\xi)) - 
x_p q(x_p) \mathcal{J}_v(\bk,\xi,X(\xi))\right],
\end{align}
where $X_0$ corresponds to the initial condition for the BK evolution of the target. The functions $\mathcal{J}$ and $\mathcal{J}_v$ can be written as Fourier transforms of coordinate-space integrals:
\begin{equation}
\hspace*{-0.7cm}
\mathcal{J}(\bk,\xi,X(\xi)) =
\int\dif^2 \br\, \rme^{-\rmi \bk \cdot \br}
J(\br,\xi,X(\xi)) \,, \quad \mathcal{J}_v(\bk,\xi,X(\xi)) =
\int\dif^2 \br\, \rme^{-\rmi \bk \cdot \br}
J_v(\br,\xi,X(\xi)) \,,
\end{equation}
where the expressions for $J$ and $J_v$ can be found in~\cite{Ducloue:2017dit}. When the transverse momentum of the produced particle is significantly larger than the target's saturation scale, it cannot be provided by the target and thus it must be balanced by the one of the unobserved gluon. Formulated in coordinate space, this constraint means that the cross-section should receive contributions from the region $x_{\perp} \sim r_{\perp}$, where $\bx$ is the transverse coordinate of the gluon. Indeed, if we consider a fixed coupling or a momentum space running coupling, we find that the contribution from the region $x_{\perp} \gg r_{\perp}$ is independent of $\br$ and thus does not contribute to the cross-section because of the final Fourier transform:
\begin{equation}
\label{jxggr}
\mathcal{J}(\bk,\xi) \sim
\frac{\abar}{2\pi^2}  
\int \dif^2 \br \,
\rme^{-\rmi \bk \cdot \br}
\underbrace{\int_{r_{\perp}^2} \frac{\dif^2 \bx }{\bx^2}\,
\left[ S((1-\xi) \bx) - S(-\xi\bx) S(\bx)
\right]}_{\br-{\rm independent}}=0 	
\quad \mbox{\rm for} \;\; x_{\perp} \gg r_{\perp} \,,
\end{equation}
and similarly for $\mathcal{J}_v$. On the contrary, if the coupling is made to depend on the parent dipole size $\br$, this will lead to a large unphysical contribution from this region. In particular, this is the case for the two running coupling prescriptions most commonly used when solving the BK equation: the smallest dipole prescription and the Balitsky prescription~\cite{Balitsky:2006wa}. Indeed, when the parent dipole is much smaller than the two daughter ones they both reduce to $\abar(r_{\perp})$.

Because of this, and the fact that the BK equation can be written using the same integrals $J$ and $J_v$, one could wonder why similar issues don't appear when solving it with these prescriptions. The reason is that the BK equation involves the difference between $J$ and $J_v$, i.e.
\begin{equation}
\mathcal{J}(\bk,\xi=1) - \mathcal{J}_v(\bk,\xi=1) 
= \int \dif^2 \br\, 
\abar(r_{\perp})\,
\rme^{-\rmi \bk \cdot \br}
\int \frac{\dif^2 \bx}{(2\pi)^2}
\frac{\br^2}{\bx^2(\bx + \br)^2}
\left[S(-\bx) S(\br + \bx) - S(\br) \right].
\end{equation}
Thus the spurious contributions coming from the large daughter dipoles region cancel in this case.

\begin{figure*}[t]
\begin{center}
\begin{minipage}[b]{0.47\textwidth}
\begin{center}
\includegraphics[width=0.95\textwidth,angle=0]{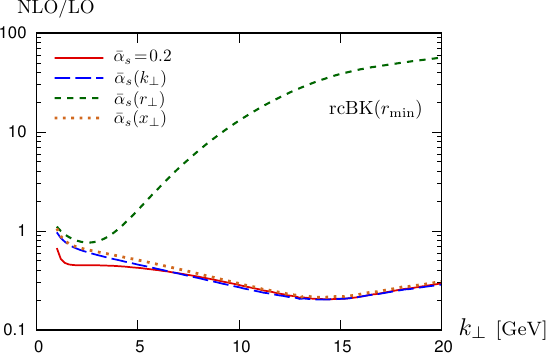}\\(A)\vspace{0cm}
\end{center}
\end{minipage}
\begin{minipage}[b]{0.47\textwidth}
\begin{center}
\includegraphics[width=0.95\textwidth,angle=0]{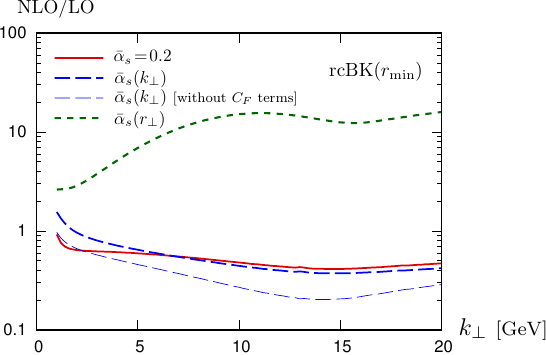}\\(B)\vspace{0cm}
\end{center}
\end{minipage}
\end{center}
\caption{\label{fig:NLOratio} Left: Ratio of the NLO multiplicity (including only the $N_c$ terms) and the LO one for different prescriptions of the running coupling. Right: Ratio of the total NLO quark multiplicity (including both the $N_c$ and $C_F$ terms) and the LO one for three running coupling prescriptions. For comparison, we also show the results for $\abar(k_{\perp})$ when including only the $N_c$ terms (same as the curve ``$\abar(k_\perp)$'' in the left panel). For both figures $\sqrt{s}=500~$GeV, $\eta=3.2$ and the evolution of the color dipoles is obtained by solving the Balitsky-Kovchegov equation with the smallest dipole prescription using an MV~\cite{McLerran:1993ni} initial condition at $X_0=0.01$.}
\end{figure*}

Based on this, we expect that using the daughter dipole prescription in the impact factor should lead to physical results: since $\abar(x_{\perp})$ does not depend on $\br$, the integral~(\ref{jxggr}) still vanishes after taking the Fourier transform when $x_{\perp} \gg r_{\perp}$. On the other hand, in the contributing region $x_{\perp} \sim r_{\perp}$, one recovers the prescription $\abar(r_{\perp})$. To illustrate this we show in Fig.~\ref{fig:NLOratio}~(A) the results for the NLO/LO ratio obtained with different prescriptions for the running coupling used in the impact factor. We observe that the results obtained with the daughter dipole prescription are close to the ones obtained with a momentum space prescription $\abar(k_{\perp})$ or with a fixed coupling. With the daughter dipole prescription it becomes possible to use the same coupling in the whole calculation, and, as in the case of a fixed coupling, there is no ambiguity between the ``subtracted'' and ``unsubtracted''~\cite{Iancu:2016vyg} formulations of the cross-section. Note, however, that it is not very natural to use the daughter dipole prescription when solving the BK equation as one generally expects the scale of the running coupling to be set by the hardest scale in the problem.

So far we considered only the NLO corrections to the cross-section proportional to $N_c$. Another source of such corrections is proportional to $C_F$. These terms are affected by the same large daughter dipoles problem as the $N_c$ terms, but an additional complication appears here. Indeed, the $C_F$ terms contain collinear divergences which have to be absorbed into the DGLAP evolution of the parton distribution functions and fragmentation functions. Because this subtraction is performed in momentum space, it is not possible to rewrite all the $C_F$ terms as double integrals over $\br$ and $\bx$. Therefore, we cannot use the daughter dipole prescription for these terms. In addition, while with a fixed or momentum space running coupling the $C_F$ terms vanish when $\xi \to 1$, this is no longer the case when the coupling depends on transverse coordinates, and this generates a spurious longitudinal logarithm. We thus consider that the most physical choice for these terms is the momentum space prescription $\abar(k_{\perp})$. In Fig.~\ref{fig:NLOratio}~(B) we show the results we obtain when including both the $N_c$ and $C_F$ NLO corrections with fixed, momentum and coordinate space running coupling. For comparison we also show the results obtained with the momentum space prescription including only the $N_c$ NLO terms. This allows us to see that the inclusion of the $C_F$ terms has a sizable effect and, being opposite in sign compared to the $N_c$ terms, they reduce the size of the NLO corrections to the cross-section.

\section{Conclusions}

In this work we have identified the origin of the troublesome results obtained with the coordinate space running coupling prescription used in~\cite{Ducloue:2017mpb} in the calculation of forward particle production at next-to-leading order. This problem is due to the fact that the Fourier transform and the choice of the running coupling prescription do not commute, and that without a careful choice of this prescription some unphysical contributions no longer cancel after the final Fourier transform. We proposed~\cite{Ducloue:2017dit} to overcome this problem by using the daughter dipole prescription for the $N_c$ terms, as this preserves the required cancellations and leads to very similar results compared to a momentum scale choice. However, the same choice cannot be made for the $C_F$ terms due to the subtraction of the collinear divergences, and the choice $\abar(k_{\perp})$ seems to be mandatory for these terms.

\section*{Acknowledgments}
This work used computing resources from CSC -- IT Center for Science in Espoo, Finland.
The work of T.L. is supported by the Academy of Finland, projects 273464 and 303756.
The work of T.L. and B.D. has been supported in part by the European Research Council, grant ERC-2015-CoG-681707.
The work of B.D., E.I. and G.S. is supported in part by the Agence Nationale de la Recherche project ANR-16-CE31-0019-01.
The work of A.H.M.
is supported in part by the U.S. Department of Energy Grant \#DE-FG02-92ER40699.












\end{document}